\documentclass[sigconf]{acmart}

\usepackage{booktabs}
\usepackage{tabularx}
\usepackage{array}
\newcolumntype{M}[1]{>{\centering\arraybackslash}m{#1}}
\newcolumntype{Y}{>{\centering\arraybackslash}X}

\usepackage{hyperref}

\usepackage{graphicx}
\graphicspath{ {figs/} }

\usepackage{xcolor}

\usepackage{amsmath}

\usepackage{pgfplots,pgfplotstable,capt-of}
\pgfplotsset{compat=1.18}
\usepgfplotslibrary{external}
\newcommand{\dataset}{FACTors}
\newcommand{\lucene}{{\tt Lucene}}

\copyrightyear{2025}
\acmYear{2025}
\setcopyright{acmlicensed}\acmConference[SIGIR '25]{Proceedings of the 48th International ACM SIGIR Conference on Research and Development in Information Retrieval}{July 13--18, 2025}{Padua, Italy}
\acmBooktitle{Proceedings of the 48th International ACM SIGIR Conference on Research and Development in Information Retrieval (SIGIR '25), July 13--18, 2025, Padua, Italy}
\acmDOI{10.1145/3726302.3730339}
\acmISBN{979-8-4007-1592-1/2025/07}
\begin{document}


\title{\dataset: A New Dataset for Studying the Fact-checking Ecosystem}
% \titlenote{Enes Altuncu is the corresponding author, with the remaining authors listed in lexicographic order of their surnames.}

\author{Enes Altuncu}
\orcid{0000-0002-1362-9700}
\affiliation{%
  \institution{University of Kent}
  \department{Institute of Cyber Security for Society (iCSS) \& School of Computing}
  \city{Canterbury}
  \state{Kent}
  \country{United Kingdom}
}
\email{drenesaltuncu@gmail.com}
\authornote{Enes Altuncu is the corresponding author; the other authors are listed in lexicographic order by surnames.}

\author{Can Ba\c{s}kent}
\affiliation{%
  \institution{Middlesex University}
  \department{Department of Computer Science}
  \city{London}
  \state{}
  \country{United Kingdom}
}
\email{c.baskent@mdx.ac.uk}

\author{Sanjay Bhattacherjee}
\affiliation{%
  \institution{University of Kent}
  \department{Institute of Cyber Security for Society (iCSS) \& School of Computing}
  \city{Canterbury}
  \state{Kent}
  \country{United Kingdom}
}
\email{s.bhattacherjee@kent.ac.uk}

\author{Shujun Li}
\orcid{0000-0001-5628-7328}
\affiliation{%
  \institution{University of Kent}
  \department{Institute of Cyber Security for Society (iCSS) \& School of Computing}
  \city{Canterbury}
  \state{Kent}
  \country{United Kingdom}
}
\email{s.j.li@kent.ac.uk}

\author{Dwaipayan Roy}
\affiliation{%
  \institution{Indian Institute of Science Education and Research Kolkata (IISER Kolkata)}
  \department{Computational and Data Sciences}
  \state{West Bengal}
  \country{India}
}
\email{dwaipayan.roy@iiserkol.ac.in}

\renewcommand{\shortauthors}{Enes Altuncu, Can Ba\c{s}kent, Sanjay Bhattacherjee, Shujun Li \& Dwaipayan Roy}

\begin{abstract}
Our fight against false information is spearheaded by fact-checkers. They investigate the veracity of claims and document their findings as fact-checking reports. With the rapid increase in the amount of false information circulating online, the use of automation in fact-checking processes aims to strengthen this ecosystem by enhancing scalability. Datasets containing fact-checked claims play a key role in developing such automated solutions. However, to the best of our knowledge, there is no fact-checking dataset at the ecosystem level, covering claims from a sufficiently long period of time and sourced from a wide range of actors reflecting the entire ecosystem that admittedly follows widely-accepted codes and principles of fact-checking.

We present a new dataset \dataset, the first to fill this gap by presenting ecosystem-level data on fact-checking. It contains 118,112 claims from 117,993 fact-checking reports in English (co-)authored by 1,953 individuals and published during the period of 1995-2025 by 39 fact-checking organisations that are active signatories of the IFCN (International Fact-Checking Network) and/or EFCSN (European Fact-Checking Standards Network). It contains 7,327 overlapping claims investigated by multiple fact-checking organisations, corresponding to 2,977 unique claims. It allows to conduct new ecosystem-level studies of the fact-checkers (organisations and individuals).

To demonstrate the usefulness of our dataset, we present three example applications. They include a first-of-its-kind statistical analysis of the fact-checking ecosystem, examining the political inclinations of the fact-checking organisations, and attempting to assign a credibility score to each organisation based on the findings of the statistical analysis and political leanings. Our methods for constructing \dataset\ are generic and can be used to maintain a live dataset that can be updated dynamically.
\end{abstract}

\begin{CCSXML}
<ccs2012>
   <concept>
       <concept_id>10002951.10003260.10003277.10003279</concept_id>
       <concept_desc>Information systems~Data extraction and integration</concept_desc>
       <concept_significance>500</concept_significance>
   </concept>
   <concept>
        <concept_id>10002951.10003260.10003261.10003270</concept_id>
        <concept_desc>Information systems~Social recommendation</concept_desc>
        <concept_significance>300</concept_significance>
        </concept>
 </ccs2012>
\end{CCSXML}

\ccsdesc[500]{Information systems~Data extraction and integration}
\ccsdesc[300]{Information systems~Social recommendation}

\keywords{False information, misinformation, disinformation, fact-checking, dataset, resources, ecosystem}


\maketitle

\section{Introduction}
%\label{sec:intro}

False and misleading information can cause serious harm to the society~\cite{tran2020}. According to the \emph{Global Risks Report 2025} from the World Economic Forum~\cite{wef2025}, information disorder is expected to become the most severe social threat in the next two years. Fact-checking organisations worldwide play a vital role in our fight against this threat. They check the veracity of suspected claims circulating online and provide verdicts. The fact-checking process has three main steps -- identifying the claim that needs investigation, evidence retrieval for assessing the claim's veracity, and veracity assessment of the claim based on the retrieved evidence. At the end of the fact-checking process, a veracity verdict is generated and a \textit{fact-checking report} is often produced to explain the verdict.

The standardisation of the fact-checking procedures is crucial for fact-checking organisations to enhance their trustworthiness and credibility. With this respect, fact-checking codes and standards, such as the International Fact-checking Network (IFCN) Code of Principles\footnote{\url{https://ifcncodeofprinciples.poynter.org/}} and the European Code of Standards for Independent Fact-Checking Organisations\footnote{\url{https://efcsn.com/code-of-standards/}}, have been implemented to promote non-partisanship, fairness, transparency (of sources, funding, organisation, and methodology), standardisation of the processes, accountability and trustworthiness towards unbiased fact-checking.

Considering the excessive amount of misinformation and disinformation circulating online, typical fact-checking processes need to be improved in several aspects, including scalability, effectiveness, and efficiency. This motivates research efforts on developing automated fact-checking solutions. With the help of ongoing advancements in artificial intelligence (AI) and natural language processing (NLP) models, various fact-checking tasks are being automated~\cite{guo2022}. Increasing efforts have been made to improve the effectiveness and efficiency of all the three steps of fact-checking, i.e., claim detection and matching~\cite{cheema2025}, evidence retrieval~\cite{samarinas-etal-2021-improving} and veracity assessment~\cite{garcialozano2020}. Developing automated fact-checking solutions (especially using machine learning) require relevant datasets for training, validation and testing purposes. Many datasets have been developed for such purposes (see Table~\ref{tab:comparison_datasets} for some examples). However, existing datasets have several limitations. Perhaps the most significant limitation is their restricted coverage, thus introducing biases such as the following two types:
\begin{itemize}
\item \textbf{Temporal bias}: Most datasets cover claims/verdicts from a specific time period, overlooking the historical context or the evolution of the claims and misinformation patterns over time.

\item \textbf{Source selection bias}: They mostly cover only a small subset of available fact-checking sources, introducing a selection bias. Further, geographic and linguistic diversity is also limited in most of them.
\end{itemize}

Many datasets contain synthetic claims/verdicts rather from the real-world~\cite{sarrouti2021}. While synthetic claims/verdicts can be useful for controlled experiments and initial model development, they often fail to capture the complexity and nuances of real-world claims/verdicts, such as the following:
\begin{enumerate}
\item \textbf{Linguistic patterns}: They may not reflect the natural language patterns and rhetorical devices commonly used in real-world misinformation.

\item \textbf{Context dependency}: They often lack the rich contextual elements that characterise real-world misinformation.

\item \textbf{Temporal dynamics}: They cannot typically capture the evolving nature of misinformation narratives over time in the real world.
\end{enumerate}

Some datasets include \textit{overlapping claims}. These are generally popular claims that draw enough interest to be investigated by multiple fact-checkers. The fact-checking process arguably becomes more robust for such claims because multiple fact-checkers investigate the same claim to provide their ``independent'' verdicts. However, an overlapping claim may have similar or dissimilar verdicts from different fact-checkers. Automating the process of arriving at a verdict on such a claim would require aggregation of these verdicts. Current approaches of aggregating verdicts on overlapping claims typically rely on \textit{simple majority voting}~\cite{akhtar2022, setty2024}. This is a simplistic approach whereby all fact-checkers are treated equally. It makes several problematic assumptions such as the following:
\begin{enumerate}
\item \textbf{Equal credibility}: It implicitly or explicitly assumes all fact-checkers possess equal credibility, expertise and confidence level.

\item \textbf{No political bias}: Fact-checkers may exhibit political biases that influence their verdicts~\cite{singh2024}, which is not considered.

\item \textbf{Equal quality}: The thoroughness of evidences used and other such quality factors can vary significantly between the fact-checking reports, which is also not considered.
\end{enumerate}
The main reason behind this simplification is that existing datasets cannot facilitate more robust methods of aggregation like weighted majority voting, where each fact-checker would be assigned a different weight~\cite{BBL2022}. The first attempt in using different credibility scores of fact-checkers for aggregating verdicts had to use synthetic data~\cite{amri2024}.

In summary, to the best of our knowledge, there is no dataset that contains fact-checking ecosystem-level data for a sufficiently long period of time, that explicitly abides by popularly agreed fact-checking codes and principles. The one that comes closest~\cite{quelle2025} is temporally biased (covering more recent years only). As a result, it is not possible to analyse the ecosystem across different fact-checkers over a long period. In addition, there is a general lack of datasets that allow us to understand many aspects of the fact-checking ecosystem, e.g., characterising the fact-checkers (both at organisational and individual levels) in terms of their experience, credibility, political inclinations, practices like rigour and completeness, etc.

To address these limitations, we introduce \emph{\dataset}, a new English fact-checking dataset with 118,112 fact-checks collected from 39 fact-checking organisations, all active members of the IFCN and/or EFCSN which declared that they publish content in English. The dataset contains real-world claims with their corresponding verdicts, as well as titles, authors, timestamps, and URLs of the fact-checking reports. It includes the \lucene\ inverted index to enable fast and efficient searching\footnote{\url{https://lucene.apache.org/core/}}. Overlapping claims fact-checked by different fact-checking organisations are presented without aggregation. Among other things, this will allow more advanced and domain-specific approaches for verdict aggregation. We also provide three examples of using the dataset to show its potential applications. To summarise, our overall contributions are as follows:

\begin{enumerate}
\item We present the first fact-checking dataset containing all reports published in English on the websites of 39 currently active IFCN/EFCSN signatories. This is almost the entire ecosystem of fact-checkers who publish reports in English while following broadly agreed codes and principles of fact-checking. Since our dataset contains all-time data, we believe it is a representative dataset for analysing the fact-checking ecosystem in several aspects, such as the consistency between the verdicts of different fact-checkers and the characteristics of fact-checks during notable events.

\item 
We normalise the verdicts to make them uniform across organisations. We also identify overlapping claims (i.e., semantically identical claims) in our dataset. We do not aggregate the verdicts of overlapping claims so that their variation can be studied further to gain deeper insights about the ecosystem.

\item As a major example application of our dataset, we share all-time statistics of the 39 fact-checking organisations and the 1,953 individuals that have written reports for them. These statistics can be used for credibility assessment of fact-checking organisations and individuals. Such credibility scores can lead to better techniques for aggregating veracity verdicts, than the commonly used simple majority voting.

\item Beyond the statistical analysis, we demonstrate the usefulness of the dataset through two other example applications. We use a pre-trained BERT-based model to detect political biases of the 39 organisations, to present a consolidated picture of the political leanings within the fact-checking ecosystem. We further assign simple credibility scores to the organisations based on their statistics.
\end{enumerate}

The remainder of the paper is structured as follows. Section~\ref{sec:related-work} provides a review of the relevant literature, with particular focus on existing fact-checking datasets and their applications in automated fact-checking, fact-checking ecosystem analysis, and credibility assessment of the fact-checkers. Section~\ref{sec:methodology} introduces the new \dataset\ dataset, detailing the data collection process, preprocessing steps, and key features that differentiate it from previous datasets. In Section~\ref{sec:example_applications}, we present example applications where our dataset can be leveraged, demonstrating its potential utility for various fact-checking-related tasks. We also discuss preliminary analyses and insights derived from the dataset. Finally, Section~\ref{sec:conclusion} summarises our findings, highlights the contributions of \dataset, and outlines potential future directions of research in which \dataset\ can be employed.

The dataset, structured as a Lucene index, along with the source code for running baseline models across various tasks, is openly available in the project repository\footnote{\url{https://github.com/altuncu/FACTors}} to support research and ensure reproducibility.

\section{Related Work}
\label{sec:related-work}

Existing fact-checking datasets have been widely used for various tasks, including misinformation detection~\cite{kwon2023}, claim verification~\cite{layne2024}, and credibility analysis~\cite{lee2023}. However, they have faced criticism for their limitations as described earlier. In addition, they do not capture ambiguous fact-checking scenarios frequently encountered in the real world. \citet{glockner2024} demonstrated that most existing datasets primarily focus on clear-cut cases where evidence either definitively supports or refutes a claim. In contrast, real-world fact-checking often involves incomplete, contradictory, or inconclusive evidence, requiring more nuanced decision-making. This oversimplification hinders the development of automated fact-checking systems that can effectively handle complex cases.

Given these challenges, our work focuses on three key areas where fact-checking datasets are most applicable: (1) automated fact-checking, which involves developing models trained on fact-checking reports to verify claims; (2) analysing the fact-checking ecosystem, which examines fact-checking practices, inconsistencies, and trends across different organisations; and (3) credibility assessment, which seeks to evaluate the trustworthiness of fact-checkers and their verdicts. We believe that \dataset\ is well-suited to these tasks, offering a more comprehensive and structured resource for advancing research in these domains. In this section, we summarise key past research work in these three subdomains, providing the context for how \dataset\ contributes to ongoing research in fact-checking. Table~\ref{tab:comparison_datasets} presents a comparative overview of relevant datasets, highlighting their characteristics in relation to \dataset.

% In addition to the shortcomings of existing datasets mentioned earlier, there have been some other criticisms like their inability to effectively include ambiguous scenarios common in real-world fact-checking. \citet{glockner2024} demonstrated that existing datasets predominantly focus on clear-cut cases where the evidence either definitively supports or refutes a claim. However, real-world fact-checking often involves situations where the evidence available is incomplete, contradictory, or insufficient to make a definitive judgment. This oversimplification in current datasets fails to prepare automated systems for the nuanced decision-making required in practical applications. We believe that \dataset\ can be leveraged in different tasks related to fact-checking, having the potential to enhance previously constructed datasets for these tasks. With this respect, this section elaborates on previous datasets that are most relevant to \dataset, and Table~\ref{tab:comparison_datasets} shows a summary of the comparison between \dataset\ and other relevant datasets. 

\begin{table*}
\centering
\small
\caption{Comparison of key existing datasets with~\dataset. The `Task' column shows three different fact-checking tasks: AFC (automated fact-checking), EA (analysing the fact-checking ecosystem), and CA (credibility assessment).}
\label{tab:comparison_datasets}
\begin{tabularx}{\linewidth}{YcccccYc}
\toprule
\textbf{Dataset} & \textbf{Year} & \textbf{Size} & \textbf{Language(s)} & \textbf{Period} & \textbf{Task} & \textbf{Source} & \textbf{Overlapping}\\
& & & & & & & \textbf{Claims}\\
\midrule
\citet{popat2016} & 2016 & 4,856 & English & Until 2016 & CA & Snopes & --\\
\citet{lim2018} & 2018 & 4,856 & English & 2013--2016 & EA & Fact Checker and Politifact & \checkmark\\
\citet{hanselowski2019} & 2019 & 6,422 & English & Unspecified & AFC & Snopes & --\\
X-FACT~\cite{gupta2021} & 2021 & 31,189 & 25 languages & Unspecified & AFC & 85 fact-checking orgs & --\\
\citet{jiang2021} & 2021 & 13,696 & English & 1994--2021 & EA & Snopes & --\\
WatClaimCheck~\cite{khan2022} & 2022 & 33,721 & English & 1996--2021 & AFC & 8 fact-checking orgs & --\\
\citet{lee2023} & 2023 & 24,169 & English & 2016--2022 & EA & 4 fact-checking orgs & \checkmark\\
\citet{lelo2023} & 2023 & 600 & Portuguese & 2022 & EA & 5 Brazilian fact-checking orgs & \checkmark\\
\citet{fernandez2023} & 2023 & 313 & Spanish & 2018--2019 & CA & Newtral & --\\
\citet{quelle2025} & 2024 & 264,487 & 95 languages & 2018--2024 & EA & Over 100 fact-checking orgs & \checkmark\\
MCFEND~\cite{li2024} & 2024 & 23,789 & Chinese & 2014--2023 & AFC & 14 fact-checking orgs & --\\
\citet{gangopadhyay2024} & 2024 & 65,121 & English & 1996--2022 & EA & ClaimsKG & \checkmark\\
\midrule
\textbf{\dataset} & \textbf{2025} & \textbf{118,112} & \textbf{English} & 1995--2025 & \textbf{EA+CA} & \textbf{39 fact-checking orgs} & \textbf{\checkmark} \\
\bottomrule
\end{tabularx}
\end{table*}

\subsection{Automated Fact-Checking}

Fact-checking reports are useful data sources for training automated fact-checking models. Therefore, previous studies commonly utilise them to obtain real-world claims and construct datasets. For example, \citet{hanselowski2019} proposed a dataset which contains 6,422 claims in English with five veracity labels. The data was collected from Snopes and covers evidence with its stance or verdict. X-FACT~\cite{gupta2021} contains 31,189 non-English claims in 25 languages with seven veracity ratings. Another example is WatClaimCheck~\cite{khan2022}, an English dataset with a size of 33,721, focusing on claim inference in automated fact-checking. The data was collected from eight fact-checking organisations, and the original verdicts were mapped to \textit{true}, \textit{false}, and \textit{partially true/false}. More recently, MCFEND~\cite{li2024} was constructed as a multi-source Chinese fact-checking dataset with a size of 23,789. It covers fact-checks collected from 14 fact-checking organisations, including Chinese articles and Chinese translations of English articles. The original ratings were mapped to \textit{real} and \textit{fake}.

Unlike existing automated fact-checking datasets focusing on organisations adopting rating scales, \dataset\ has a more comprehensive coverage of fact-checking organisations regardless of how they shape their verdicts. This enables the development of a more robust and comprehensive dataset without overlooking a considerable amount of fact-checked claims currently available.

\subsection{Analysing the Fact-Checking Ecosystem}

Fact-checking organisations are relatively new and not as firmly established as traditional news outlets. Consequently, researchers have sought to analyse the fact-checking ecosystem to better understand its characteristics and to identify ways to improve fact-checking methodologies. For example, \citet{jiang2021} collected 13,696 fact-checking reports from Snopes between 1994 and 2021. They explored ten different types of false information derived by clustering the collected articles and analysed how each type has evolved from before 2010 to the end of 2020, focusing on notable events. More recently, \citet{kumar2024} qualitatively analysed the declared methodologies of seven Indian fact-checking organisations in terms of transparency. They concluded that fact-checking was conducted by the studied organisations in a transparent and systematic manner. In another recent study, \citet{gangopadhyay2024} analysed the fact-checks released in English covered by the ClaimsKG knowledge graph, including 65,121 claims investigated by seven fact-checking organisations between 1996 and 2022, to explore the characteristics of the existing fact-checking landscape and to analyse the underlying biases. They also identified 2,450 similar claims investigated by multiple organisations by considering cosine similarities between the sentence embeddings generated from the claims using the sentence transformer model Sentence-BERT (SBERT)~\cite{reimers2019}. Finally, \citet{quelle2025} constructed a multilingual dataset of 264,487 fact-checks in 95 languages, published between 2018 and 2024, to explore the prevalence and dynamics of online false information. They also utilised language-agnostic BERT sentence embedding (LaBSE) with cosine similarity to identify multiple investigated claims, and with a similarity threshold of 0.875, they observed 10.26\% of all claims being fact-checked more than once.

As the number of fact-checking organisations continues to grow, the likelihood of multiple organisations investigating the same claims has increased. This motivated some researchers to analyse the consistency between different fact-checking organisations' fact-checks. To exemplify, \citet{lim2018} analysed 1,503 fact-checks from two US fact-checking organisations, Politifact and Fact Checker, published between 2013 and 2016. Their results showed that only 77 fact-checks overlap, with 49 of those fact-checks having verdicts that were agreed upon by both organisations. Expanding on this, \citet{lee2023} examined 22,349 fact-checks from Snopes and Politifact in terms of the consistency of their verdicts. Among 749 matching claims, they identified 228 cases (6.5\%) where verdicts diverged and only one conflicting verdict after accounting for differences in rating scales. The study also analysed 1,820 fact-checks from the UK-based fact-checking company Logically and Australian Associated Press FactCheck in terms of their fact-checking behaviours. Similarly, \citet{lelo2023} analysed 600 fact-checks released by five Brazilian fact-checking organisations during the 2022 presidential elections in Brazil. Out of these fact-checks, they identified 142 overlapping fact-checks, 97 of which with an agreed verdict.

\subsection{Credibility Assessment}

Credibility assessment is a crucial task that can help the development of more effective solutions for detecting false information. While previous credibility assessment methods largely focused on information sources and content credibilities, the credibility assessment of fact-checkers is understudied~\cite{srba2024}. Although fact-checking organisations are generally regarded as highly credible in existing ratings (e.g., Media Bias/Fact Check\footnote{\url{https://mediabiasfactcheck.com/}}), there is still a need to differentiate between organisations and individual fact-checkers based on their previous fact-checking performances, especially when there is a lack of consensus among different fact-checkers on the veracity of a claim. As an example of research efforts in this domain, \citet{popat2016} used data collected from Snopes for a case study aiming to assess its credibility based on language and web-source reliability features. The dataset involves 4,856 claims until 2016, labelled as true or false. More recently, \citet{fernandez2023} analysed ideological bias -- an indicator of credibility -- using 313 fact-checks from the Spanish organisation Newtral, supplemented with interviews of its fact-checkers. Their findings suggest that fact-checking reflects journalistic decisions rather than systematic bias due to the flexible application of verification protocols.

Building on these efforts, \citet{amri2024} proposed AFCC, a system incorporating mathematical definitions for fact-checker credibility assessment and consensus inference. Instead of relying on majority voting, AFCC assigns weights to fact-checkers based on their past performance and calculates consensus using a weighted average of ratings for the same claim. The system was evaluated through theoretical analysis and simulations under three different scenarios involving varying proportions of biased and unbiased fact-checkers. However, AFCC is yet to be tested with real-world fact-checking datasets, highlighting the need for data-driven validation.

Although further research is needed on the credibility assessment of fact-checkers, meaningful progress requires ecosystem-level datasets that provide comprehensive insights into the broader fact-checking landscape. Such datasets should capture historical fact-checking data, including patterns in verdicts, fact-checker reliability, and potential biases. We believe that \dataset\ can address this gap by offering a structured resource for analysing the credibility of fact-checkers at scale.

\section{\dataset: A New Dataset of Fact-Checks} 
\label{sec:methodology}

To bridge the research gaps highlighted earlier, we present a new dataset, \dataset, designed to support various fact-checking-related tasks, including automated fact-checking, fact-checking ecosystem analysis, and credibility assessment. \dataset\ consists of 118,112 fact-checks published in English by 39 fact-checking organisations, along with detailed metadata that provides valuable insights into the fact-checking landscape.

Tables~\ref{tab:dataset-description} and \ref{tab:example-row} outline the structure of the dataset, describing its columns and presenting an example row to illustrate the available information. Our dataset captures essential attributes such as the claim, its verdict, publication date, the fact-checker (organisation and individual), and additional metadata that can be leveraged for various analytical and machine-learning applications. The following subsections provide a comprehensive overview of the dataset construction process, including data collection methodologies, preprocessing steps, and considerations made to ensure the quality and reliability of the dataset.

\begin{table}[!htb]
\centering
\caption{Description of our dataset \dataset, highlighting its unique features compared to existing fact-checking datasets.}
\label{tab:dataset-description}
\begin{tabularx}{\linewidth}{l|X}
\toprule
\textbf{Field Name} & \textbf{Description}\\
\midrule
Row ID & Primary Key\\
Report ID & ID given to each unique report\\
Claim ID & ID given to each unique claim\\
Claim & Textual claim being fact-checked\\
\textit{Content} & \textit{(not published to prevent copyright infringement)}\\
Date published & Date of publication of the report\\
Author & Author(s) of the fact-checking report\\
Organisation & Name of the fact-checking organisation publishing the report\\
Original verdict & Conclusion of the fact-check\\
Title & Heading of the report\\
URL & Online link to the report\\
Normalised rating & One of six predefined ratings derived from the original verdict\\
\bottomrule
\end{tabularx}
\end{table}

\begin{table}[!htb]
\centering
\caption{An example of one entry of fact taken from our dataset \dataset.}
\label{tab:example-row}
\begin{tabularx}{\linewidth}{l|X}
\toprule
\textbf{Field Name} & \textbf{Description}\\
\midrule
Row ID & 83122\\
Report ID & 43638\\
Claim ID & 80198\\
Claim & HAARP is responsible for recent floods in Spain, New Mexico, and elsewhere.\\
\textit{Content} & \textit{(not published to prevent copyright infringement)}\\
Date published & 2024-11-08T16:11:56\\
Author & Rahul Rao\\
Organisation & Science Feedback\\
Original verdict & Incorrect\\
Title & No, HAARP can't create floods, climate change can make heavy rainfall more extreme - Science Feedback\\
URL & \url{https://science.feedback.org/review/haarp-cant-create-floods-climate-change-can-make-heavy-rainfall-more-extreme/}\\
Normalised rating & False\\
\bottomrule
\end{tabularx}
\end{table}

\subsection{Data Collection}

To construct a robust and representative dataset, we first identified all fact-checking organisations which were verified signatories of the International Fact-checking Network (IFCN) Code of Principles\footnote{\url{https://ifcncodeofprinciples.poynter.org/}} and the European Code of Standards for Independent Fact-Checking Organisations\footnote{\url{https://efcsn.com/code-of-standards/}} (as of November 2024). In addition, we only considered the organisations with English appearing as one of their primary languages on their respective IFCN and/or EFCSN descriptions to provide adequate samples for each covered organisation. This amounted to 42 fact-checking organisations, with 34 solely verified by IFCN, 1 verified by EFCSN only, and 7 verified by both IFCN and EFCSN.

Data was collected from these organisations through web scraping. For the implementation of the automated scripts used to crawl the website of each organisation -- referred to as \emph{spiders} -- we employed the Python library Scrapy\footnote{\url{https://scrapy.org}}. Additionally, we utilised the \texttt{scrapy-playwright}\footnote{\url{https://github.com/scrapy-plugins/scrapy-playwright}} plugin to integrate the Playwright\footnote{\url{https://playwright.dev/}} web browser automation framework with Scrapy in circumstances when the webpages to be scraped required user input, e.g., scrolling down to load more content on dynamic webpages. For each fact-checking organisation covered, we implemented a spider using the relevant CSS selectors as input, corresponding to the locations of various types of information on the webpage to be scraped. They include report-level information, (i.e., title, author, date, content, fact-checked claim, and final verdict) as well as domain-level information (i.e., the list of fact-checking reports and the button for accessing the next page of the list). In addition, we considered ClaimReview\footnote{\url{https://www.claimreviewproject.com/}} structured data placed in the HTML codes of the scraped webpages whenever it was available. Claims and verdicts extracted from ClaimReview data were given priority over those obtained from the unstructured webpage content. However, we observed that some fact-checking organisations neither included ClaimReview data nor showed claim-verdict pairs in an identifiable format on their website. For such cases, we leveraged the title and description meta tags, which generally include the investigated claims and the corresponding verdicts. Depending on the data presentation practices of each organisation, we determined whether the title or description was more suitable as the claim or verdict.

The scraping was performed between January 16 and February 8, 2025. During scraping, we paid special attention to conduct the task ethically and responsibly by several means, including obeying the rules defined in the \texttt{robots.txt} file and throttling the crawling speed. Consequently, we were able to crawl around 140k fact-checking reports from 39 organisations. However, data from the remaining three organisations (Reuters, The Washington Post, and Deutsche Welle) could not be collected due to specific challenges -- anti-scraping measures, paywalls, and server connection issues, respectively. We decided to not include them in our dataset.

\subsection{Data Cleaning and Preparation}

After data collection, we cleaned the initial dataset in several steps. While we only considered fact-checking reports published in English, we noticed that the collected data contained some non-English reports mistakenly included on the English websites of some organisations publishing in multiple languages. Therefore, we first identified non-English reports using the Python library \texttt{langdetect}\footnote{\url{https://github.com/Mimino666/langdetect}} and excluded them from the dataset. Then, we removed the rows with one or more missing key values, such as title, content, claim, or verdict, as well as those with too short claim and content fields containing fewer than 10 characters. Furthermore, the dates in the dataset were standardised to the ISO 8601\footnote{\url{https://www.iso.org/iso-8601-date-and-time-format.html}} format.

We used the titles and descriptions provided in meta tags as some claims and verdicts in our dataset. So they are likely to contain some redundant phrases which are not part of the claims or verdicts, such as ``Fact-check: '' or ``Here's the truth''. By employing the word tokeniser module of NLTK~\cite{bird2006} and reviewing the most common 500 $n$-grams ($3 \leq n \leq 6$), we identified 105 such phrases and removed them. Moreover, we deduplicated the dataset by leveraging sentence vector embeddings generated using SBERT~\cite{reimers2019}. The claims fact-checked by the same organisation and having the embeddings with a cosine similarity higher than 0.95 were considered duplicates that were removed\footnote{These decisions were made based on a thorough manual review of the selected claims.}.

After the cleaning procedure, we prepared the dataset for release. To ensure that the dataset does not violate copyrights, we removed the full-text reports from the public dataset. We converted our dataset into a \lucene\ inverted index to enable fast and effective searching. In addition, we assigned ID numbers to each unique row of the dataset and each report, and placed them in separate columns in the dataset.

\subsection{Verdict Normalisation}

Fact-checking organisations have diverse ways to format the verdicts they reach upon investigating claims. Therefore, it is a common practice for fact-checking datasets to convert the original verdicts to a predefined set of simplified ratings to facilitate the usage of datasets in automation. Nevertheless, most existing datasets only cover data from organisations that implemented a rating scale with distinct truth levels, for the sake of simplicity. This simplifies the need for verdict normalisation to a keyword mapping for such datasets, which is not the case for \dataset. Therefore, we followed a three-step methodology to normalise the original verdicts into a six-point rating scale --- \emph{true}, \emph{partially true}, \emph{false}, \emph{misleading}, \emph{unverifiable}, and \emph{other}. Firstly, we manually reviewed all unique verdicts shorter than five words to identify the original verdicts that can conveniently be mapped to one of the six predefined ratings. This corresponded to 68 original verdicts, covering 72,309 fact-checks of 33 organisations in our dataset. In the next step, we used the content-verdict pairs of this subset to fine-tune a base RoBERTa~\cite{liu2019} model and assigned the predicted labels as the normalised verdicts to the remaining fact-checks. The model was trained with a learning rate of 3e-5 over three epochs, resulting in an accuracy of 0.849 with a train/test split ratio of 90:10, which seems promising when previous research is considered~\cite{jiang2020, woloszyn2021}. Finally, we went through the predictions by the model with a confidence level below 0.5, corresponding to 1,564 labels, and manually corrected the predictions when needed.

\subsection{Identifying Semantically Identical Claims}

As a key feature of our dataset, we identified overlapping claims -- semantically identical claims fact-checked by multiple organisations -- in our dataset. Similar to our deduplication approach, we leveraged the cosine similarity of sentence embeddings generated using SBERT. To find the optimum cosine similarity threshold, we first sorted the claim pairs according to their cosine similarity and manually observed them for different cosine similarity levels. We noticed that the likeliness of finding overlapping claims was considerably higher when the cosine similarity is at least 0.75, so we excluded the pairs with a cosine similarity less than 0.75. For a more fine-grained analysis, we randomly sampled 1,000 claim pairs with a uniform distribution of the cosine similarity and annotated each pair as overlapping or not. Then, we looked at the cosine similarity threshold value that would have resulted in 95\% precision in our sample, which appeared as 0.88. By applying this threshold to all claim pairs in the entire dataset, we obtained 7,327 fact-checks with overlapping claims, referring to 2,977 unique claims. To express the overlapping claims, we added a separate column, \emph{claim\_id}, to our dataset where the overlapping claims share the same claim ID. We did not check the overlapping claims manually, so there are likely false positives among them, as well as false negatives that we were unable to identify.

\subsection{\dataset\ and Its Features}

Our final dataset consists of 118,112 fact-checks performed by 39 fact-checking organisations. Considering the reports covering multiple fact-checked claims, \dataset\ involves 117,993 unique fact-checking reports. When it comes to the number of claims covered, our dataset contains 113,762 unique claims, 2,977 of which were investigated by multiple fact-checking organisations. The average word count for the full-text reports is 537,22. Ultimately, the covered fact-checking reports were written by 1,953 different authors. The percentage of fact-checking reports with one or more authors' names provided is 80.87\%, while 1.36\% of the reports were written by multiple authors.

We release our dataset as a CSV file along with the associated source codes and the results of the conducted analyses based on the dataset as a GitHub repo at \url{https://github.com/altuncu/FACTors}. Moreover, to facilitate immediate use by the research community, we also make \dataset\ freely available as a comprehensive Apache Lucene (version 8.11.0) index\footnote{The link to the index is available in the GitHub repo.} that encapsulates all dataset information. The index stores the complete metadata for each fact-checking report, including claims, verdicts, publication details and author information. This distribution approach enables researchers to directly query and analyse the data without additional preprocessing or index construction steps. The Lucene index provides fast lookup capabilities through its inverted index structure, allowing efficient searching across all fields with support for both exact matching and fuzzy searches. Researchers can easily retrieve fact-checks by specific organisations, authors or time periods. Moreover, the index supports complex boolean queries and relevance-based ranking, enabling sophisticated analysis of fact-checking patterns. For researchers preferring Python-based text processing, tools like Pyserini~\cite{Lin_etal_SIGIR2021_Pyserini} can be used to interact with the index, providing a familiar interface for information retrieval and analysis tasks.

\section{Example Applications on Downstream Tasks}
\label{sec:example_applications}

As we believe \dataset\ can be utilised in several tasks related to fact-checking, particularly analysing the characteristics of the ecosystem and credibility assessment, this section is intended to shed light on some potential applications of our dataset.

\subsection{Statistical Analysis of Fact-Checking Ecosystem and Fact-Checkers}
\label{subsec:statistics}

Collecting all historical data during the construction of \dataset\ enabled us to perform different statistical analyses based on the collected data which can help better understand the characteristics of the fact-checking ecosystem and fact-checkers.

\begin{figure}[!ht]
\centering
\includegraphics[width=\linewidth]{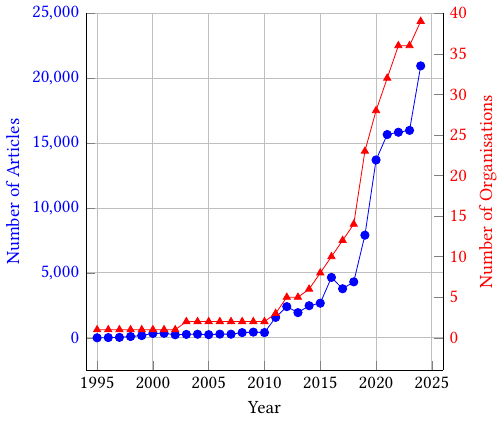}
\caption{Yearly statistics showing the number of fact-checking organisations publishing fact-checks per year (\textcolor{red}{red}) and the total number of fact-checking reports released annually (\textcolor{blue}{blue}).}
\label{fig:yearly-stats}
\end{figure}

To begin with, in our dataset covering data from 39 fact-checking organisations, the number of those that released at least one fact-check per year has dramatically increased after 2010, as shown in Figure~\ref{fig:yearly-stats}, consistent with earlier findings~\cite{cherubini2016}. Concordantly, the total number of fact-checking reports released per year began climbing after 2010 and surged after the 2016 US presidential elections, reaching up to over 20,000 annually. This indicates how rapidly the amount of false and misleading information has increased with an emphasis on the need for more fact-checking organisations and more scalable fact-checking solutions.

\newlength\figwidth
\setlength{\figwidth}{0.9\linewidth}

Deep-diving into the characteristics of fact-checkers, both organisations and individuals, our dataset also allows for calculating some useful statistics. In this regard, we generated the statistics described below for each organisation and author (i.e., employee of a fact-checking organisation) we covered in \dataset\ to offer potential features that can be leveraged to characterise and distinguish different fact-checkers. Table~\ref{tab:stats_min_max} provides an overview of the fact-checking organisations and authors with the minimum and maximum values for various key statistics, including the duration of their fact-checking activity in days (experience), the total number of fact-checks conducted, the percentage of unique fact-checks, the average time interval between consecutive fact-checks (fact-checking rate in days), the number of contributing authors, the number of organisations they are associated with, and the average word count of their fact-checking reports. By highlighting these extreme values, the table offers insights into the variability of fact-checking practices across different organisations and authors, shedding light on differences in productivity, collaboration, and content length within the fact-checking ecosystem.

\begin{itemize}
\item \textbf{Fact-checking experience}: It is the time difference between a fact-checker's first and last published fact-checking report.

\item \textbf{Number of fact-checks}: This refers to the total number of fact-checking reports a fact-checker has published. 

\item \textbf{Percentage of unique fact-checks}: It is the proportion of fact-checked claims that have not been investigated by any other fact-checker previously. 

\item \textbf{Fact-checking rate}: It is the mean and standard deviation of how frequently a fact-checker publishes fact-checking reports.

\item \textbf{Number of authors}: It is the number of authors each organisation has employed.

\item \textbf{Number of organisations}: It is the number of fact-checking organisations each author has published with.

\item \textbf{Word count}: It is the mean and standard deviation of the number of words in the fact-checking reports published by a fact-checker.
\end{itemize}

\begin{table}
\centering
\small
\caption{Different statistics calculated for each organisation and author covered by our dataset, with the minimum and maximum values obtained.}
\label{tab:stats_min_max}
\begin{tabularx}{\linewidth}{Ycccc}
\toprule
& \multicolumn{2}{c}{\textbf{Organisations}} & \multicolumn{2}{c}{\textbf{Authors}}\\
\textbf{Statistic} & \textbf{Min} & \textbf{Max} & \textbf{Min} & \textbf{Max}\\
\midrule
Experience (days) & 32 & 10,710 & -- & 9,432\\
Number of fact-checks & 96 & 20,977 & 1 & 4,416\\
Unique fact-checks (\%) & 80.22 & 100 & 0 & 100\\
Fact-checking rate (days) & 0.046 & 18.9 & 0 & 1,902\\
Number of authors & 1 & 889 & -- & --\\
Number of organisations & -- & -- & 1 & 2\\
Word count & 214.1 & 1,778.6 & 7.57 & 5,075\\
\bottomrule
\end{tabularx}
\end{table}

\subsection{Political Bias Analysis of Fact-Checking Organisations}
\label{subsec:politicalbias}

Political bias is a crucial factor that can negatively impact the credibility of a fact-checking organisation. Therefore, as another example application of our dataset, we attempt to shed light on whether existing fact-checking organisations published reports which exhibited political bias towards left or right. For this task, we used a pre-trained BERT-based model for detecting political bias in given texts, politicalBiasBERT\footnote{\url{https://huggingface.co/bucketresearch/politicalBiasBERT}}. The model predicts the political ideology from provided texts with the probabilities of the text leaning towards left, right, and centre. We ran this model on the \emph{content} column of our dataset (not published to prevent copyright infringement) and obtained the sets of probabilities in the format [left, centre, right]. Then, we considered the value with the highest probability as the main prediction of the model and mapped the resulting value to -1, 0, or 1 for left, centre, and right, respectively. Finally, the mean and standard deviation values were calculated for each organisation.

\begin{figure}[!ht]
\centering
\includegraphics[width=\linewidth]{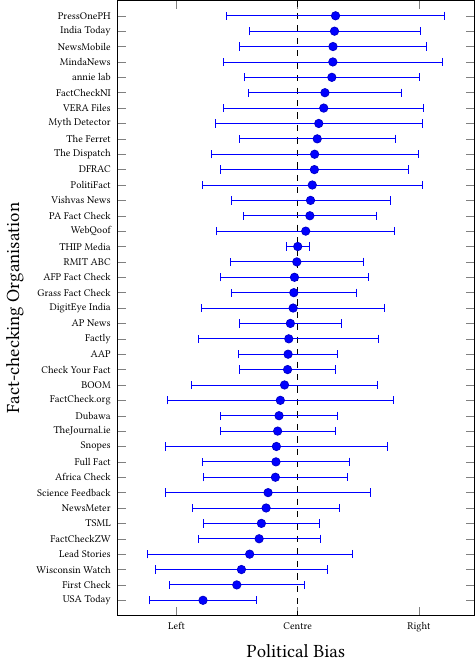}
\caption{Predicted political bias scores of fact-checking organisations.}
\label{fig:political_bias}
\end{figure}

The calculated bias scores based on the predictions are shown in Figure~\ref{fig:political_bias}. The distribution of scores suggests that fact-checking organisations are more likely to lean towards political left, yet there are a significant number of organisations exhibiting bias towards the right. While the predicted bias scores are not decisive and limited by the accuracy of the model, it implies that fact-checking organisations can also exhibit political bias in their fact-checking reports. In addition, political bias detectors might be a helpful distinguishing factor in determining the credibility of fact-checking organisations, especially when their verdicts for a claim contradict.

\subsection{Credibility Assessment of Fact-Checkers}

Sections~\ref{subsec:statistics} and~\ref{subsec:politicalbias} present several factors that can be used as different parameters in credibility assessment tasks. Therefore, this example application demonstrates how different factors derived from historical fact-checking data can be leveraged to produce credibility scores for fact-checking organisations and individuals. The produced scores can then be utilised to develop a more advanced verdict aggregation algorithm, rather than simply applying majority voting when there exist multiple (and possibly contradicting) verdicts for a claim.

Let $\text{FC}_i$ ($1\leq i\leq n$) denote one of $n$ fact-checkers and $\text{CAF}_j$ ($1\leq j\leq m$) denote one of $m$ credibility assessment factors. Then, let a predefined coefficient $c_j$ indicate the degree to which the factor $\text{CAF}_i$ impacts the assessed credibility of $\text{FC}_i$, and $e_j$ be the negativity effect of the factor $\text{CAF}_i$, which equals to 0 for positive (effective) factors and 1 for negative (ineffective) factors. Further, let $\text{Rank}(i, j)$ be a function providing the rank of a fact-checker $\text{FC}_i$ in terms of the factor $\text{CAF}_j$, among all the fact-checkers. Based on all the notations and definitions, one possible credibility score $\text{CredScore}_i$ of the fact-checker $\text{FC}_i$ can be calculated as follows:

\begin{equation}
\label{eq:credscore}
\text{CredScore}_i = \frac{1}{m} \sum_{j=1}^m \frac{(-1)^{e_j} c_j}{\text{Rank}(i, j)}
\end{equation}

Based on the above definition, we applied Eq.~\eqref{eq:credscore} to all fact-checking organisations and individuals we have in our dataset, based on the statistics calculated earlier. Since determining the weights for each factor requires further research beyond the scope of this paper, we assign $c_j=1$ for all $j$, assuming that all the factors have the same impact on credibility. Regarding the negativity effect parameter $e_j$, we simply assume the factors mentioned in Section~\ref{subsec:statistics} positively affect credibility ($e_j=0$) while political bias negatively affects credibility ($e_j=1$).

\begin{figure*}[!ht]
\centering
\includegraphics[width=\linewidth]{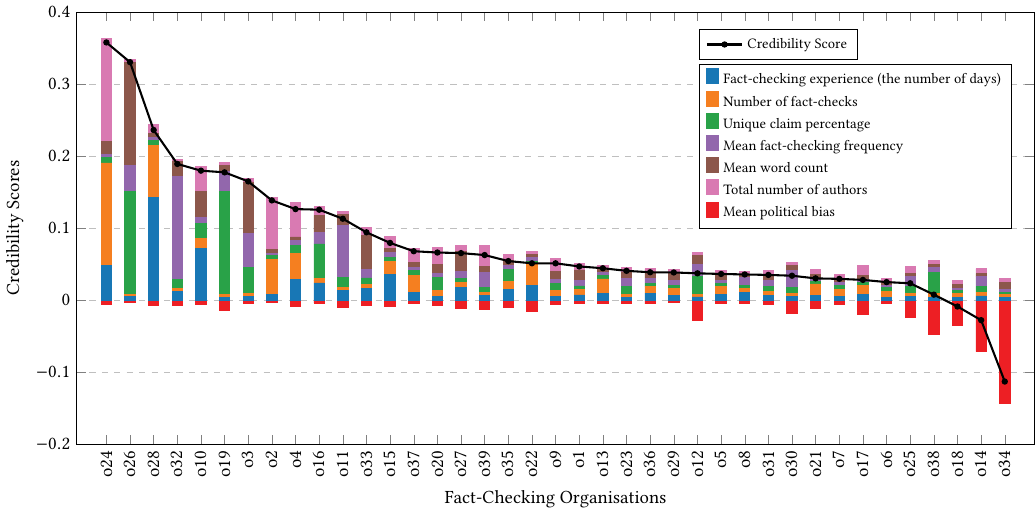}
\caption{Credibility scores of different (anonymised) fact-checking organisations, with contributions of each factor to the final score.}
\label{fig:credibility_scores}
\end{figure*}

The credibility scores calculated are shown in Figure~\ref{fig:credibility_scores}, along with the contributions of each included factor to the final score. For this experiment, we anonymised the names of the organisations to highlight that it is an example application, rather than a judgement about the credibility of fact-checking organisations which require further research. However, the distribution of the scores implies that fact-checking organisations are distinguishable in terms of their assessed credibilities, depending on the set of factors used for calculating the credibility scores.

Differentiating fact-checkers based on their assessed credibility is particularly important when a claim is investigated by multiple fact-checkers with conflicting verdicts. The assignment of credibility scores, therefore, addresses the limitations of the commonly used majority voting approach for aggregating multiple verdicts reached for the same claim. And, it enables the use of more advanced solutions, such as weighted voting games~\cite{chalkiadakis2012}, which can better reflect real-world scenarios.

\section{Conclusion}
\label{sec:conclusion}

The increasing prevalence of misinformation necessitates robust fact-checking mechanisms that can scale with the rapid dissemination of false claims. While significant progress has been made in developing automated fact-checking systems, existing datasets present several challenges. Our analysis of key fact-checking datasets has highlighted these limitations and provided insights into potential improvements for future dataset development. In this paper, we introduced \dataset, a comprehensive fact-checking dataset containing 118,112 fact-checks from 39 IFCN and/or EFCSN member organisations. Our dataset addresses several critical limitations of existing fact-checking datasets by preserving individual organisational verdicts for overlapping claims, maintaining complete report metadata, and providing efficient access through a Lucene-based index structure. The comprehensive temporal coverage and organisational breadth of \dataset\ enable ecosystem-level analyses that were previously challenging or impossible to conduct.

Beyond practical applications, the proposed dataset enables theoretical explorations using logical and game-theoretical methods. Graded verdicts facilitate concepts such as voting games, equilibria, and non-classical logics~\cite{BBL2022}, offering a structured approach to misinformation analysis. These methods can enrich discussions on misinformation, disinformation, and higher-order misinformation~\cite{harris} while introducing epistemic and strategic perspectives. Thus, as part of the novel application of the datasets, future work should explore these theoretical dimensions, leveraging interdisciplinary research to enhance fact-checking methodologies and the broader fight against misinformation. Notably, such explorations are not feasible with any existing datasets, underscoring the need for a more comprehensive and structured resource to support these advanced analytical approaches.

Our work has some limitations that we would like to address in the future. Initially, our dataset only covers reports written in English. Therefore, enhancing \dataset\ with multilingual data can enable to present a worldwide picture of the fact-checking ecosystem. Besides, our verdict normalisation is bounded by the accuracy of the fine-tuned model we used due to the existence of non-structured verdicts in our dataset. In addition, the identification of overlapping claims may contain false positives and may have missed false negatives, therefore, manually verifying all such claims may be necessary and developing a more accurate overlapping claim detection method can be useful. Finally, we were unable to collect data from three fact-checking organisations among those we initially identified due to some challenges we mentioned in the paper.

\begin{acks}
The authors would like to thank the anonymous reviewers for their careful review of the work.
\end{acks}

\bibliographystyle{ACM-Reference-Format}
\balance
\bibliography{main}

\end{document}